\documentclass[12pt]{article}
\usepackage{epsfig}

\def\beq{\begin{equation}}
\def\eeq#1{\label{#1}\end{equation}}
\def\eeqn{\end{equation}}

\def\beqa{\begin{eqnarray}}
\def\eeqa#1{\label{#1}\end{eqnarray}}
\def\eeqan{\end{eqnarray}}

\let\bar=\overbar

\def\Dslash{\not{\hbox{\kern-4pt $D$}}}
\def\dslash{\not{\hbox{\kern-2pt $\del$}}}

\def\msb{{\bar{\ssstyle M \kern -1pt S}}}

\def\Title#1{\begin{center} {\Large {\bf #1} } \end{center}}

\begin{document}

\begin{flushright}
UdeM-GPP-TH-12-216
\end{flushright}

\Title{\boldmath Extracting Weak Phases Cleanly from Charmless 3-Body
  $B$ Decays\footnote{Talk given at the {\it 7th International
      Workshop on the CKM Unitarity Triangle}, University of
    Cincinnati, Cincinnati, Ohio, USA, October 2012. Talk based on
    work done in collaboration with M. Imbeault, N. Rey-Le Lorier and
    B. Bhattacharya.}}

\bigskip\bigskip

%+\addtocontents{toc}{{\it D. Reggiano}}
%+\label{ReggianoStart}

\begin{raggedright}  

{\it David London\index{London, D.}\\
Physique des Particules, Universit\'e de Montr\'eal\\
C.P. 6128, succ.\ centre-ville\\
Montr\'eal, QC, Canada H3C 3J7}
\bigskip\bigskip
\end{raggedright}

\begin{quote}

\centerline{Abstract}

In the past, it was believed that one cannot obtain clean weak-phase
information from the measurement of CP-violating asymmetries in 3-body
$B$ decays. Recently it was shown that this is not true -- by
expressing the decay amplitudes in terms of diagrams and using Dalitz
plots, one can resolve all the difficulties and cleanly extract weak
phases. In this talk I describe how this is done, and present
preliminary results on the measurement of $\gamma$ using the decays $B
\to K \pi \pi$ and $B \to KK{\bar K}$.

\end{quote}

The standard method for obtaining clean information about the weak
Cabibbo-Kobayashi-Maskawa (CKM) phases is through the measurement of
indirect (mixing-induced) CP-violating asymmetries in $B^0(t) \to f$.
This requires that $f$ be a CP eigenstate. Because of this, the
conventional wisdom is that one cannot obtain such clean CKM
information from 3-body decays since final states such as $K_S
\pi^+\pi^-$ are not CP eigenstates -- the value of its CP depends on
whether the relative $\pi^+\pi^-$ angular momentum is even (CP $+$) or
odd (CP $-$).

There are some exceptions. If the final state contains truly identical
particles -- e.g.\ $K_S \pi^0\pi^0$ -- it is a CP eigenstate. (Here
the relative $\pi^0\pi^0$ angular momentum is necessarily even, which
means the state is CP $+$.) Also, for $B^0\to K^+ K^- K_S$, Belle used
an isospin analysis to differentiate CP $+$ and CP $-$. They found
that it is dominantly CP $+$ \cite{BKKKS}.

Unfortunately, even for these exceptions, there is an additional
problem. The procedure for getting clean weak-phase information from
indirect CP asymmetries only works if the decay is dominated by
amplitudes with a single weak phase. However, in general these decays
receive significant contributions from amplitudes with a different
weak phase. In order to extract the weak phases, one needs a way of
dealing with this ``pollution.''

Recently it was shown that all of these difficulties can be overcome
\cite{3body1,3body2,3body3}. I describe the method below.

The first ingredient is the use of Dalitz plots. Consider the decay
$B \to P_1 P_2 P_3$ ($P$ is a pseudoscalar meson), in which each $P_i$
has momenta $p_i$. One can construct the three Mandelstam variables:
\begin{equation}
s_{12} \equiv \left( p_1 + p_2 \right)^2 ~~,~~~~
s_{13} \equiv \left( p_1 + p_3 \right)^2 ~~,~~~~
s_{23} \equiv \left( p_2 + p_3 \right)^2 ~.
\end{equation}
These are not independent, but obey the relation
\begin{equation}
s_{12} + s_{13} + s_{23} = m_B^2 + m_1^2 + m_2^2 + m_3^2 ~.
\end{equation}

The Dalitz plot is given in terms of two Mandelstam variables, say
$s_{12}$ and $s_{13}$. For the decay amplitude, we write
\begin{equation}
{\cal M}(B \to P_1 P_2 P_3) = \sum_j c_j e^{i\theta_j} F_j(s_{12},s_{13}) ~.
\end{equation}
Here the sum is over all decay modes (resonant and non-resonant).
$c_j$ and $\theta_j$ are the magnitude and phase of the $j$
contribution, relative to one of the channels. The distributions $F_j$
describe the dynamics of the individual decay amplitudes, and take
different forms for the various contributions. The key point is the
following: in the experimental Dalitz-plot analyses, explicit
expressions for the $F_j$ are assumed (e.g.\ Breit-Wigner).  Then a
maximum likelihood fit over the entire Dalitz plot gives the best
values of the $c_j$ and $\theta_j$. Thus, {\it the decay amplitude
  ${\cal M}(s_{12},s_{13})$ is known.}

With this information the CP of the final state can now be fixed. For
example, suppose the final state has CP $+$ when the amplitude is
symmetric under $P_2 \leftrightarrow P_3$ (as is the case for the
final state $K_S \pi^+\pi^-$). We can find this amplitude from the
above:
\begin{equation}
{\cal M}_{sym} = \frac{1}{\sqrt{2}} \left[ {\cal M}(s_{12},s_{13}) + {\cal M}(s_{13},s_{12}) \right] ~.
\label{symamp}
\end{equation}
Using this, it is possible to compute the $B \to P_1 P_2 P_3$
observables. E.g.\ the indirect CP asymmetry is given by
\begin{equation}
\label{S_sym}
S = {\rm Im} \left[ e^{-2i\phi_M} \, \frac{{\bar{\cal M}}_{sym}}{{\cal M}_{sym}} \right] ~.
\end{equation}

Note: all observables are momentum dependent -- they take different
values at each point in the Dalitz plot.

The second ingredient is the use of diagrams \cite{3body1}. In 2-body
decays all amplitudes are expressed in terms of color-allowed and
color-suppressed tree, gluonic penguin, and electroweak penguin (EWP)
diagrams; annihilation/exchange-type diagrams are neglected. In 3-body
decays, one has similar diagrams. Here one has to ``pop'' a quark pair
from the vacuum.  For the diagrams we add the subscript ``1'' if the
popped quark pair is between two non-spectator final-state quarks, and
``2'' if it is between two final-state quarks including the spectator.

\begin{figure}[!htbp]
	\centering
		\includegraphics[height=3.5cm]{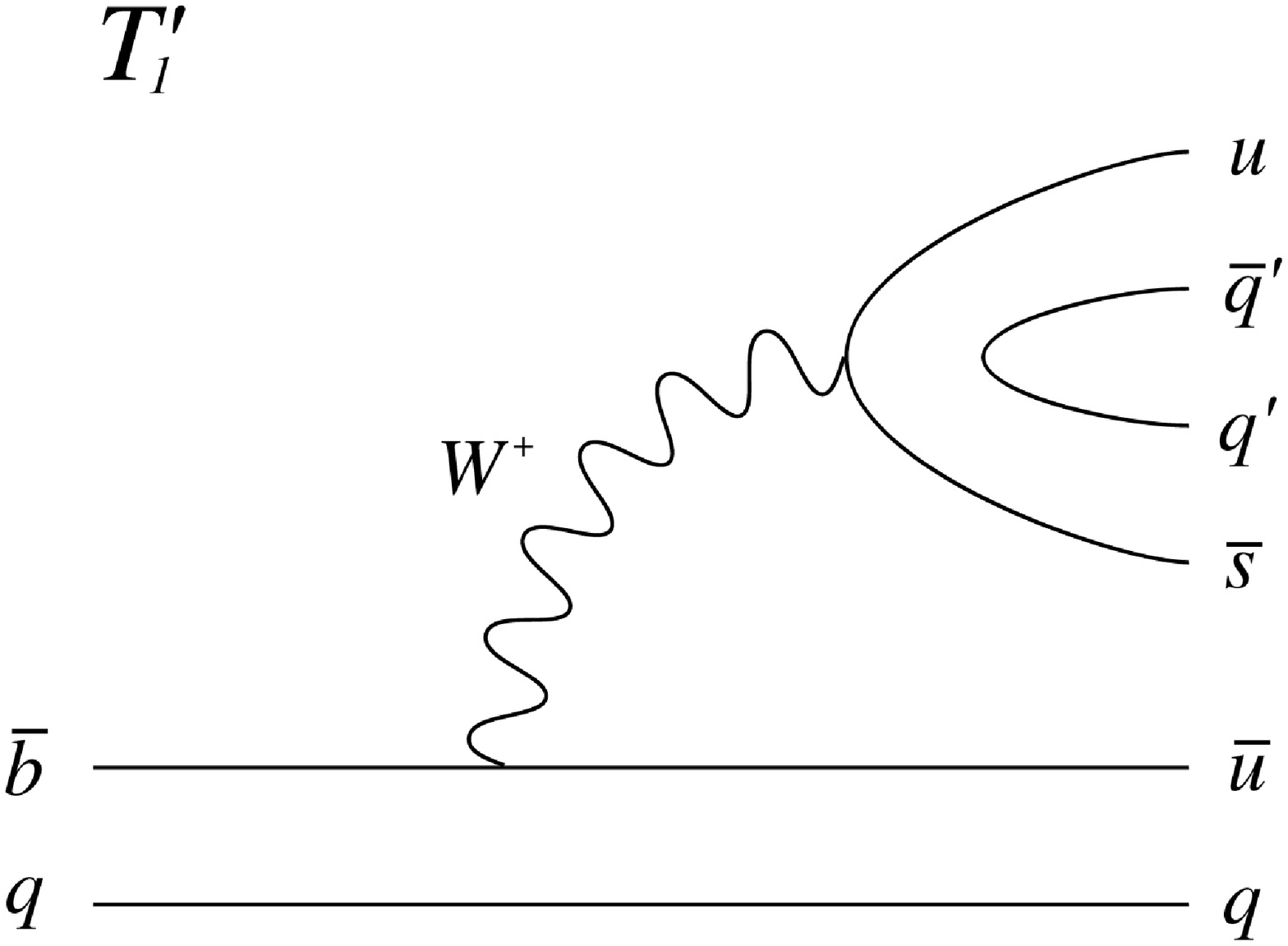}
		~~~~~~~ \includegraphics[height=3.5cm]{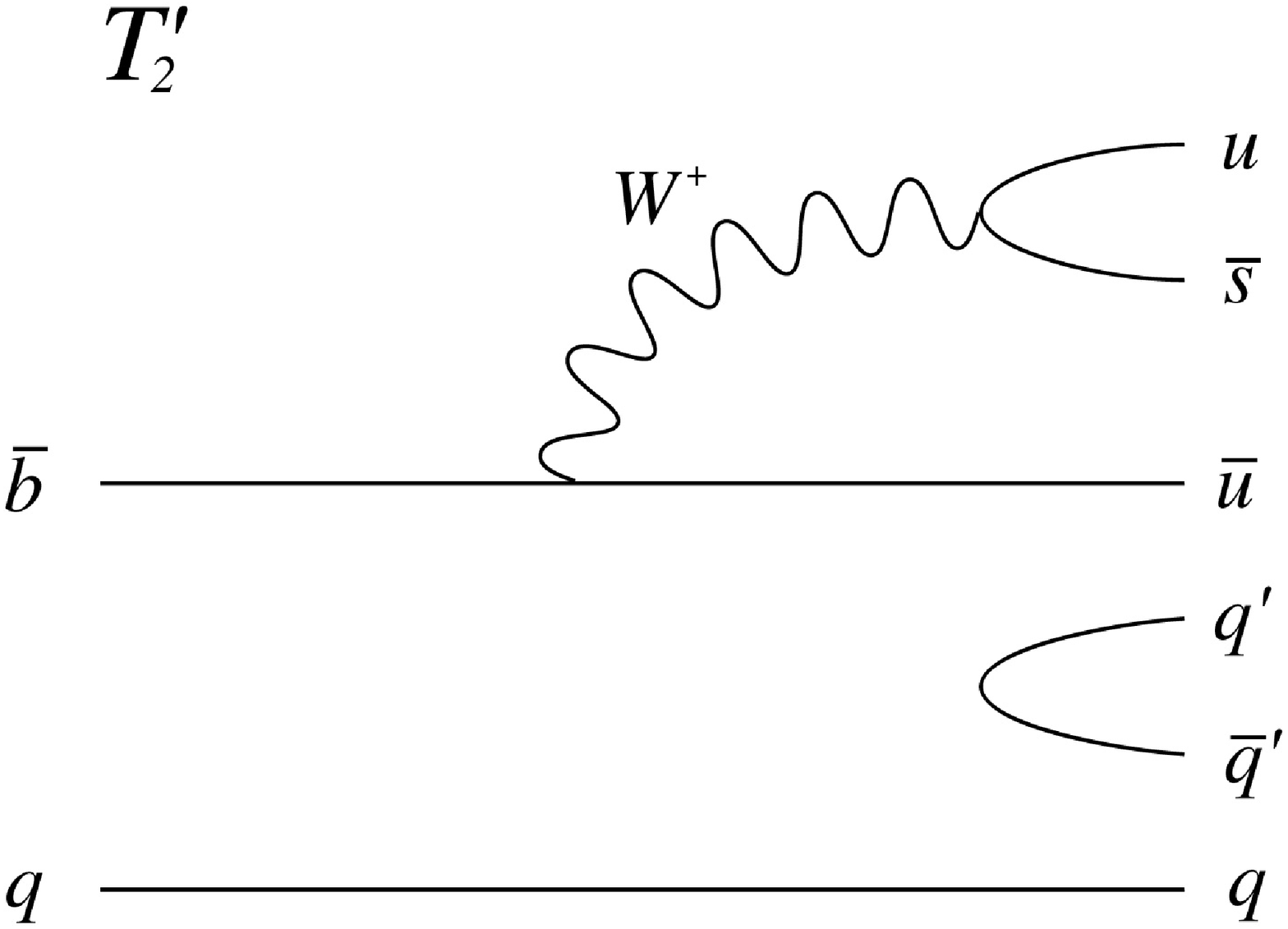}
\end{figure}
The above figure shows the $T'_1$ and $T'_2$ diagrams contributing to
$B \to K \pi \pi$ (as this is a ${\bar b} \to {\bar s}$ transition,
the diagrams are written with primes). The other diagrams ($C'_1$,
$C'_2$, $P'_1$, $P'_2$, $P'_{EW1}$, $P'_{EW2}$, $P^{\prime C}_{EW1}$,
$P^{\prime C}_{EW2}$) are obtained similarly from the 2-body diagrams.

Note: unlike the 2-body diagrams, the 3-body diagrams are momentum
dependent. This must be taken into account whenever the diagrams are
used.

Now, in $B \to K \pi$ decays there are relations between the EWP and
tree diagrams under flavor SU(3) symmetry \cite{NRGPY}. Recently it
was shown that similar EWP-tree relations hold for $B \to K \pi \pi$
decays \cite{3body2}. The Wilson coefficients obey $c_1/c_2 =
c_9/c_{10}$ to about 5\%, in which case these relations take the
simple form (the exact relations are given in Ref.~\cite{3body2})
\begin{equation}
P'_{EW1} = \kappa T'_1~,~~~
P'_{EW2} =\kappa T'_2 ~~~;~~~~~~
P^{\prime C}_{EW1} = \kappa C'_1~,~~~
P^{\prime C}_{EW2} = \kappa C'_2~,
\label{EWPtree}
\end{equation}
where
\begin{equation}
\kappa \equiv - \frac{3}{2} \frac{|\lambda_t^{(s)}|}{|\lambda_u^{(s)}|} \frac{c_9+c_{10}}{c_1+c_2} ~,
\end{equation}
with $\lambda_p^{(s)}=V^*_{pb} V_{ps}$.

However, there is an important caveat. Under SU(3), the final state in
$B \to K \pi \pi$ involves three identical particles, so that the six
permutations of these particles (the group $S_3$) must be taken into
account.  But the EWP-tree relations hold only for the totally
symmetric state. Thus, the analysis must be carried out for this
state. The fully symmetric state can be found from the Dalitz
plot. Instead of the amplitude which is symmetric only under $P_2
\leftrightarrow P_3$ [Eq.~(\ref{symamp})], we define
\begin{eqnarray}
{\cal M}_{fully~sym} & = & 
\frac{1}{\sqrt{6}} \left[ {\cal M}(s_{12},s_{13}) + {\cal M}(s_{13},s_{12}) +
  {\cal M}(s_{12},s_{23}) \right. \nonumber\\
&& \hskip1.5truecm \left. +~{\cal M}(s_{23},s_{12}) + {\cal M}(s_{23},s_{13}) + {\cal
    M}(s_{13},s_{23}) \right] ~.
\end{eqnarray}
All observables, such as the indirect CP asymmetry [see
  Eq.~(\ref{S_sym})] are computed using ${\cal M}_{fully~sym}$.

Once the full decay amplitudes are expressed in terms of diagrams, one
can perform an analysis like that done with 2-body decays -- one can
combine the amplitudes for different decays in order to isolate and
extract a CKM phase. I now give an example of such an analysis
involving $B \to K \pi \pi$ and $B \to KK{\bar K}$ decays
\cite{3body3}. (Note: SU(3) is assumed.)

There are 6 decays of the type $B^+/B^0 \to K\pi\pi$. Decays with two
$\pi^0$'s are excluded as being too difficult experimentally. Also,
$B^+ \to K^0\pi^+\pi^0$ is not independent -- its amplitude is
proportional to that of $B^0 \to K^+\pi^0\pi^-$ \cite{3body1}.  There
are therefore only three $B \to K \pi \pi$ decays to consider.

The $B \to K \pi \pi$ amplitudes in which the $\pi\pi$ pair is
symmetrized are:
\begin{eqnarray}
2 A(B^0 \to K^+\pi^0\pi^-)_{sym} &=& T'_1 e^{i\gamma}+C'_2 e^{i\gamma} - \kappa \left(T'_2 + C'_1\right) ~, \nonumber\\
\sqrt{2} A(B^0 \to K^0\pi^+\pi^-)_{sym} &=& -T'_1 e^{i\gamma}-C'_1 e^{i\gamma}-{\tilde P}'_{uc} e^{i\gamma}+ {\tilde P}'_{tc} \nonumber\\
&& \hskip1.5truecm +~\kappa \left(\frac13 T'_1 + \frac23 C'_1 - \frac13 C'_2\right) ~, \nonumber\\
\sqrt{2} A(B^+ \to K^+\pi^+\pi^-)_{sym} &=& -T'_2 e^{i\gamma}-C'_1 e^{i\gamma}-{\tilde P}'_{uc} e^{i\gamma}+ {\tilde P}'_{tc} \nonumber\\
&& \hskip1.5truecm +~\kappa \left(\frac13 T'_1 - \frac13 C'_1 + \frac23 C'_2 \right) ~.
\end{eqnarray}
These expressions hold even under the full SU(3) symmetry
\cite{3body2}.

There are four $B \to KK{\bar K}$ decays in which the final $KK$ pair
is in a symmetric isospin state. However, only the amplitudes of 
%$B^+ \to K^+ K^+ K^-$ and $B^+ \to K^+ K^0 {\bar K}^0$ are
%proportional to those of
$B^0 \to K^+ K^0 K^-$ and $B^0 \to K^0 K^0 {\bar K}^0$ are independent
\cite{3body1}.  These are
\begin{eqnarray}
\sqrt{2} A(B^0 \to K^+ K^0 K^-)_{sym} &=& -T'_2 e^{i\gamma}-C'_1 e^{i\gamma}
-{\tilde P}'_{uc} e^{i\gamma}+ {\tilde P}'_{tc} \nonumber\\
&& \hskip0.8truecm +~\kappa \left( \frac13 T'_1 - \frac13 C'_1  + \frac23
C'_2 \right) ~, \nonumber\\
A(B^0 \to K^0 K^0 {\bar K}^0)_{sym} &=& {\tilde P}'_{uc} e^{i\gamma}- {\tilde P}'_{tc} 
+ \kappa \left(\frac23 T'_1 + \frac13 C'_1 + \frac13 C'_2 \right) ~.
\end{eqnarray}
Note: since SU(3) has been assumed, $B \to KK{\bar K}$ diagrams in
which the popped quark pair is $s{\bar s}$ are equivalent to $B \to K
\pi \pi$ diagrams with a popped $u{\bar u}$ or $d{\bar d}$. This
implies that $A(B^+ \to K^+\pi^+\pi^-)_{sym} = A(B^0 \to K^+ K^0
K^-)_{sym}$.

It is straightforward to show that one can combine the diagrams into
``effective diagrams'' $T'_a$, $T'_b$, $P'_a$, $P'_b$, $C'_a$
\cite{3body3}, giving
\begin{eqnarray}
\label{effamps}
2 A(B^0 \to K^+\pi^0\pi^-)_{sym} &=& T'_a e^{i\gamma} + T'_b e^{i\gamma} - C'_a - \kappa T'_b ~, \nonumber\\
\sqrt{2} A(B^0 \to K^0\pi^+\pi^-)_{sym} &=& - T'_a e^{i\gamma} - P'_a e^{i\gamma} + P'_b ~, \\
\sqrt{2} A(B^0 \to K^+ K^0 K^-)_{sym} &=&  - P'_a e^{i\gamma} + P'_b - C'_a ~, \nonumber\\
A(B^0 \to K^0 K^0 {\bar K}^0)_{sym} &=& P'_a e^{i\gamma} - T'_b e^{i\gamma} - \frac{1}{\kappa}  C'_a e^{i\gamma} 
- P'_b + \kappa T'_a + \kappa T'_b + C'_a ~.\nonumber
\end{eqnarray}
The 5 effective diagrams involve 10 unknown theoretical parameters: 5
magnitudes of diagrams, 4 relative strong phases, and $\gamma$. But
there are 11 (momentum-dependent) experimental observables: the decay
rates and direct asymmetries for $B^0 \to K^+\pi^0\pi^-$, $B^0 \to
K^0\pi^+\pi^-$, $B^0 \to K^+ K^0 K^-$ and $B^0 \to K^0 K^0 {\bar
  K}^0$, and the indirect asymmetries of $B^0 \to K^0\pi^+\pi^-$, $B^0
\to K^+ K^0 K^-$ and $B^0 \to K^0 K^0 {\bar K}^0$.  {\it With more
  observables than theoretical parameters, $\gamma$ can be extracted
  from a fit.} Furthermore, because the observables and diagrams are
momentum dependent, this analysis applies to every point in the Dalitz
plot. Thus, this method actually constitutes many independent
measurements of $\gamma$! These can be averaged, reducing the error.

The above is a broad overview of the $K \pi \pi$/$KK{\bar K}$ method
of measuring $\gamma$. However, using the experimental Dalitz-plot
data from BaBar \cite{BaBarDP}, my collaborators and I are in the
process of carrying out this analysis. Details are given in the talk
by B. Bhattacharya \cite{bhujyo}, but here is a summary. Only 14
points in the Dalitz plots have been used in this preliminary
analysis, SU(3) breaking has not been taken into account, not all
sources of error have been included, and there are multiple
overlapping solutions (discrete ambiguities).  With these caveats, the
initial result is
\begin{equation}
\gamma = \left( 81^{+4}_{-5}~({\rm avg.}) \pm 4~({\rm std.~dev.}) \right)^\circ ~.
\end{equation}
This is consistent with independent direct measurements of
$\gamma$. The Particle Data Group gives $\gamma = \left(
66^{+11}_{-10} \right)^\circ$ \cite{pdg}.

Though preliminary, the result is extremely encouraging. It does
indeed appear that one can cleanly extract weak-phase information from
3-body $B$ decays, contrary to what was previously thought.
Furthermore, there are indications that these measurements might be of
high precision.

Of course, our analysis is based only on published data. I therefore
strongly encourage the experimentalists to incorporate this method
into their measurements of the 3-body $K \pi \pi$/$KK{\bar K}$ Dalitz
plots. There is no doubt that many of the outstanding question marks
could be better treated with a complete analysis of the experimental
data, and the error on $\gamma$ might be reduced even further.

\bigskip
This work was financially supported by NSERC of Canada.

\end{document}